\documentclass[pra,nofootinbib]{revtex4}

\usepackage{graphicx}
\usepackage{amssymb}
\usepackage{amsmath}
\usepackage{amsfonts}
\usepackage{epstopdf}
\usepackage{epsfig}
\usepackage{wrapfig}

\begin{document}

\title{Critique of Fault-Tolerant Quantum Information Processing }
\author{Robert Alicki}
\affiliation{Institute of Theoretical Physics and Astrophysics, University of
Gda\'nsk, Poland }

\begin{abstract}
This is a chapter in a book \emph{Quantum Error Correction} edited by  D. A. Lidar and   T. A. Brun,  and published by Cambridge University Press (2013)\\ (http://www.cambridge.org/us/academic/subjects/physics/quantum-physics-quantum-information-and-quantum-computation/quantum-error-correction)\\ presenting the author's view on feasibility of fault-tolerant quantum information processing.
\end{abstract}
\maketitle

\section{Introduction}
The idea of fault-tolerant quantum computation presents a bold challenge to the rather well
established principles  called the Strong Church-Turing Thesis\index{Strong Church-Turing thesis} and the\index{Bohr Correspondence Principle} Bohr Correspondence Principle.
One of the variations of the Strong Church-Turing Thesis (SCTT), which is not due to Church or Turing, but rather gradually  developed  in the field of computational complexity\index{computational complexity} theory, is the following \cite{Bernstein:1997:11}:\\

\emph{Any ``reasonable'' model of computation can be efficiently simulated on a probabilistic Turing machine.}\\ 

This can be expressed even less formally but  more practically  as: \\

\emph{No computer can be more efficient than a digital one equipped with a random number generator.}\\
Here, a computer A is ``more efficient''  than a computer B if A can solve, in  polynomial time, a problem which cannot be solved in a polynomial time\index{polynomial time!solution} by computer B.\\

On the other hand the Bohr Correspondence Principle (BCP) demands that:\\

\emph{Classical physics and quantum physics give the same answer when the system become large.}\\

A more rigorous form of the BCP is the following:\\

\emph{For large systems the experimental data are consistent with classical probabilistic models.}\\

Here, by a \emph{classical probabilistic model} we mean a model in which all observables are described by real functions, and all states by  probability distributions on a certain set (``phase-space'').\\

The conditions under which quantum and classical physics\index{Bohr Correspondence Principle} agree are referred to as the correspondence limit, or the classical limit. Bohr provided a rough prescription for the correspondence limit: it occurs when the quantum numbers describing the system are large, meaning some quantum numbers of the system are excited to a very large value. Note that the number of particles which form a physical system should also be considered a quantum number.\\
One should not expect that the  principles above can be proved in the sense of mathematical theorems. They are rather similar (and perhaps related) to the second law of thermodynamics,  which is supported by numerous theoretical models of very different levels of generality analyzed within  more or less rigorous frameworks, and a large body of experimental data.

First of all one should notice, that in practice,  
\emph{the BCP implies the SCTT}.
Indeed, in classical physics one has, essentially, two models of computers: digital and analog. The latter could in principle be more powerful because they use continuous variables described by real numbers. However, the finite accuracy of state preparation
and measurement combined with the chaotic behavior of generic classical dynamics implies that analog computers cannot out perform digital ones. The only known resource of Nature which remains is ''quantumness,'' i.e., superposition\index{superposition}s and entanglement of quantum states. However, all complexity notions are asymptotic, valid in the limit of large input and consequently large computer. Therefore, if the BCP is universally valid, a large quantum computer is equivalent to a classical analog machine.

An often used heuristic argument for the robustness of quantum information processing relies on the linearity of the Schr\"odinger equation, in contrast to nonlinearity of the evolution equation for a generic classical system. This is, however, a misunderstanding as both theories should be compared on the level of dynamical equations for states or observables, which are linear in both the quantum and classical cases. Moreover, there exists a large body of evidence that on a logarithmic\index{logarithmic time scale} (with respect to Hilbert space dimension) time scale quantum dynamics follows its corresponding semiclassical limit with the same sensitivity to external perturbations \cite{Zurek:1994:2508,Alicki:2001:Oxford}.

In the next sections we discuss two challenges to the BCP related to quantum information processing:  threshold theorems\index{theorem!threshold} for Fault-Tolerant Quantum Computation (FTQC), and the idea of quantum memory\index{quantum memory} based on topological degrees of freedom. In contrast to often used phenomenological approaches, we scrutinize the basic assumptions using first-principle Hamiltonian models. In the case of FTQC we argue that the basic assumptions of the phenomenological models disagree with the fundamental features of the Hamiltonian approach. Concerning topological quantum information processing, although some interesting phenomena are observed, e.g., for\index{Kitaev model!four-dimensional} 4D-Kitaev models, it seems unlikely that all the desired properties of a quantum memory are achievable.

\section{Fault-Tolerant Quantum Computation}
The most important results of the theory of quantum error correction\index{quantum error correction} and FTQC are the threshold theorems \cite{Aharonov:2008:1207}
Here we reproduce a less formal presentation of this theory following the review article \cite{Knill:2002:188}.\\

\textbf{Theorem 1.} \emph{Assume the requirements for scalable quantum information processing (see below). If the error per gate is less than a threshold, then it is possible to efficiently quantum compute arbitrarily accurately.}\\

The assumptions of the above theorem are the following.\\
\emph{ a) Scalability :} The systems must be able to support any number of qubits.\\
\emph{ b) State preparation and measurement}. One must be able to prepare any qubit in a standard initial state with probability $1-\epsilon$ and measure any qubit in the logical basis
with  accuracy  $1-\epsilon$ at the end of computation, where $\epsilon$ is a small number.\\ 
\emph{c) Quantum control:} One must be able to implement a universal set of unitary gates\index{universal!set of quantum gates} acting on a small number of qubits (typically one and two). A certain amount of parallelism in gate application is also required.\\
\emph{ d) Errors:} The error probability per gate must be below a threshold and satisfy certain independence and\index{locality} locality
 properties.\\

All these conditions present formidable technological challenges to experimentalists and engineers. From the point of view of fundamental physical principles the last condition concerning the properties of noise seems to be the most important and also the most questionable one. In the following, commonly used phenomenological models of noise will be compared with first-principle ones.

\subsection{Phenomenological vs. Hamiltonian models of FTQC}

We begin with the phenomenological model of quantum computation as presented, for example, in \cite{Aharonov:2008:1207}. One  assumes that the quantum computer (QC) consists of $N= N_r + N_a$ qubits where $N_r$ belongs to the \emph{register} and 
$N_a$ form an ancillary system used in the error correction procedures. The difference between register and ancillas\index{ancillary qubits} becomes important only at the end of the computation when the information is extracted only from register qubits. One treats now the whole computer as an open system with the Hilbert space ${\cal H}_N$. Gates are maps acting on the density matrices $\rho$ as
${\cal U}\rho = U\rho U^{\dagger}$ where $U$ is a unitary matrix from the universal set of gates (say, 1 or 2 qubit gates).
One  divides the computation time into time steps and $l(k)$ denotes a \emph{location}, i.e., space-time coordinates of the qubits participating in the same gate (including trivial gates) at  time step $k$. The influence of the noise\index{noise} is described by a map ${\cal E}_k$ acting between  time steps $k-1$ and $k$. The  assumptions concerning the \emph{error maps} ${\cal E}_k$ that allow one to prove threshold theorems\index{theorem!threshold} are the following:\\

A1) ${\cal E}_k$ is linear .\\

A2) The error map can be always written as  
$
{\cal E}_k = I + \sum_{L} \Phi_L
$
where $\Phi_L$ is a linear map acting only on qubits from a subset $L$ containing $|L|$ qubits.
There exists a constant $\eta$,  called the \emph{error per gate}, and an overall constant $C$
such that  $\|\Phi_L\| \leq C\eta^{|L|}$, where the norm\index{norm} is an appropriate one for superoperators, such as the diamond norm.\\

In the literature it is usually assumed that the error maps are completely positive. This is not necessary, as in the proofs 
only linearity and the norm estimates are used, and it is even not desirable as shown below.
 
We now compare  the phenomenological model above with the standard description in terms of the reduced dynamics of an open quantum system  \cite{Alicki:2007:Springer,Breuer:2002:OxfordUniversityPress}. One can treat  the whole computer as an open system with the Hilbert space ${\cal H}_N$ weakly interacting with a bath described by the Hilbert
space  ${\cal H}_B$. The dynamics of the total system is governed by the Hamiltonian
\begin{equation}
H(t) = H'_Q(t) + H_B + {\lambda} H_{int},
\label{g1:hamopen}
\end{equation}
where $H'_Q(t)$ describes a \emph{bare} time-dependent control over the quantum computer (QC), $H_B$ is the bath Hamiltonian, and $\lambda$ is the coupling constant of the QC-bath interaction Hamiltonian $H_{int}$. The bare Hamiltonian\index{Hamiltonian!bare}\index{Hamiltonian!physical} differs from the \emph{physical Hamiltonian} $H_Q(t)$ by the presence of (generally frequency cut-off dependent) counterterms, which compensate for Hamiltonian corrections due to the interaction with an environment\index{Lamb shift}\index{Stark shift} (i.e., terms compensating for the Lamb and Stark shift). In the following we assume that the physical Hamiltonian can be perfectly designed and implemented. In all formulas below we use a renormalized picture, the evolution is always governed by the physical (renormalized) Hamiltonian, and all Hamiltonian corrections are removed.\\
\par
A standard assumption is that the initial state is a  product state $\rho(0)\otimes \rho_B$, and that $[H_B,\rho_B] =0$,
which is consistent with the\index{weak coupling regime} weak coupling regime. This leads to the following expression for the reduced dynamics of the QC
\begin{equation}
\rho(t) = {\cal E}(t)\rho(0)= {\rm Tr}_B\bigl(U(t)\rho(0)\otimes\rho_B U^{\dagger}(t)\bigr)\ ,
\label{g1:reddyn}
\end{equation}
where
\begin{equation}
\frac{dU(t)}{dt}= -iH(t)U(t)\ ,\  U(0)= I\ .
\label{g1:reddyn1}
\end{equation}
Provided ${\cal E}(t)^{-1}$ exists, which is the generic case, one can always treat $\rho(t)$ as the solution of the following \emph {time-convolutionless master equation}\index{master equation!time-convolutionless} \cite{Breuer:2002:OxfordUniversityPress}
\begin{equation}
\frac{d}{dt}\rho(t) = {\cal L}(t)\rho(t)\ ,\ {\cal L}(t)=\Bigl(\frac{d}{dt}{\cal E}(t)\Bigr){\cal E}(t)^{-1}\ .
\label{g1:conv}
\end{equation}
The following notation for unitary and nonunitary\index{superpropagator} \emph{superpropagators}, defined in terms of ordered exponentials, will be used:
\begin{equation}
{\cal U}(t_2, t_1) = {T_+}\exp\Bigl\{ -i\int_{t_1}^{t_2}{\cal H}_Q (t)dt\Bigr\}\ ,\  {\cal H}_Q(t)\rho \equiv[H_Q(t),\rho] 
\label{g1:prop}
\end{equation}
\begin{equation}
{\cal E}(t_2, t_1) = {T_+}\exp\Bigl\{ \int_{t_1}^{t_2}{\cal L} (t)dt\Bigr\}\ .
\label{g1:prop1}
\end{equation}
It is usually assumed that the QC works according to a clock with time step $\tau$, such that for the total time of computation
$t= K\tau$ we can represent the dynamical map ${\cal E} (K\tau)$ as a product of unitary and nonunitary maps
\begin{equation}
{\cal E}(K\tau)= {\cal E}_K{\cal U}_K\cdots {\cal E}_2{\cal U}_2{\cal E}_1{\cal U}_1\ .
\label{g1:dynprod}
\end{equation}
where
\begin{equation}
{\cal U}_k = {\cal U}(t_k, t_{k-1})\ ,\ {\cal E}_k= {\cal E}(t_k, t_{k-1}){\cal U}_k^{-1}\ .
\label{g1:dyngate}
\end{equation}
The implementation of the quantum algorithm in terms of gates is performed in such a way that for a given computation step $k$
the unitary superoperator \index{superoperator!unitary} ${\cal U}_k$ can be decomposed into a product of commuting (say one or two-qubit)  superoperators\index{superoperators!commuting}
corresponding to disjoint \emph{locations} $l(k)$ which involve all qubits of the QC
\begin{equation}
{\cal U}_k = \prod_{l(k)} {\cal U}_{l(k)} \ ,\ {\cal U}_{l(k)} = {T_+}\exp\Bigl\{ -i\int_{t_{k-1}}^{t_k}{\cal H}_{l(k)}(t)\,dt\Bigr\}
\label{g1:loc}
\end{equation}
where  $H_{l(k)}(t)$ is a one or two-qubit Hamiltonian implementing the gate. 
\par
We discuss now the properties of error maps defined by Eqs. (\ref{g1:prop1}) and (\ref{g1:dyngate}). Although the total evolution map ${\cal E} (t)$ is completely positive, the error maps ${\cal E}_k$ need not be, since the joint system-bath state may be non-classically correlated at the time instants $t_k$. They are obviously linear , i.e., satisfy A1 but in general there is no reason to assume the validity of the condition A2. This follows from the fact that the structure of the noise map depends not only on the interaction Hamiltonian, which  usually has a local structure, but also on the system Hamiltonian $H_Q(t)$. 
Assuming that the interaction Hamiltonian
is given by  a single-qubit coupling to the bath
\begin{equation}
H_{int}= \sum_{\mu = x,y,z}\sum_{j=1}^N\sigma^{\mu}_j\otimes B^{\mu}_j \equiv\sum_{\alpha }\sigma^{\alpha}\otimes B_{\alpha} ,
\label{g1:int}
\end{equation}
we can derive the approximate form of the error map ${\cal E}_k$ in the lowest order\index{Born approximation} (Born) approximation. We begin with the definition of reduced dynamics in the interaction picture with respect to the free dynamics ${\cal U}_0(t)$ generated by the Hamiltonian $H_0(t)= H_Q(t) + H_B $:
\begin{equation}
{\cal E}^{int}(t)\rho= {\rm Tr}_B\bigl(({\cal U}_0(t))^{-1}{\cal U}(t)\rho\otimes\rho_B \bigr)={T_+}\exp\Bigl\{ \int_{0}^{t}{\cal L}^{int}(s)\,ds\Bigr\}\ .
\label{g1:reddynint}
\end{equation}
Applying van Kampen's cumulant expansion technique \cite{Breuer:2002:OxfordUniversityPress}, one can write ${\cal L}^{int}(t) = \sum_n \lambda^n {\cal L}^{int}_n(t)$ and explicitly compute the leading (Born) term as
\begin{equation}
{\cal L}^{int}_2(t)\rho= -\lambda^2\int_0^t ds\, {\rm Tr}_B [H_{int}(t),[H_{int}(s),\rho\otimes\rho_B]] \ .
\label{g1:Bornint}
\end{equation}
Comparing the definitions (\ref{g1:dynprod}), (\ref{g1:dyngate}) and (\ref{g1:reddynint}), one can show that the error map\index{error!map} can be expressed as
\begin{equation}
{\cal E}_k= {\cal U}(t_k, t_{k-1})\,{T_+}\exp\Bigl\{ \int_{t_{k-1}}^{t_k}{\cal L}^{int}(t)\,dt\Bigr\}\,{\cal U}^{-1}(t_k, t_{k-1})\ .
\label{g1:dyngate1}
\end{equation}
Combining (\ref{g1:int}), (\ref{g1:reddynint}),(\ref{g1:Bornint}), and (\ref{g1:dyngate1}), one obtains the structure of the leading terms of the error map 
\begin{eqnarray}
{\cal E}_k &=& I + \lambda^2\sum_{\alpha\beta}\int_{t_{k-1}}^{t_k} dt\int_0^t ds F_{\alpha\beta}(t-s) 
\sigma^{\alpha}\cdot \sigma^{\beta}(t-s)  
\nonumber\\
&+& {\rm similar\  terms}\Bigr)
 + {\cal O}(\lambda^4),
\label{g1:error}
\end{eqnarray}
with the correlation function\index{correlation function}
\begin{equation}
 F_{\alpha\beta}(t)= {\rm Tr}_B\bigl(\rho_B B_{\alpha}(t)B_{\beta}\bigr) \ .
\label{g1:corr}
\end{equation}

Note that only in the case of Dirac-delta correlations\index{correlations!Dirac-delta} of the reservoir, i.e., $F_{\alpha \beta}(t-s) = \delta_{\alpha \beta}(t-s)$, does the single-qubit coupling to the bath produce  single-qubit error maps to leading order. Otherwise, $\sigma^{\beta}(t-s)$  contains multi-qubit contributions to the noise ``coming from the past'' \cite{Alicki:2002:062101}.
The Hamiltonian
$H_Q(t)$ should allow coupling between all qubits of the QC, for otherwise the system could be decomposed into completely independent components. Therefore, one can expect that the number of qubits which may contribute to the noise operator $\sigma^{\beta}(\tau)$  grows roughly linearly with $\tau$. According to Eq. (\ref{g1:error}), the weight of terms which contain  $n$-qubit operators with $n\sim \tau$ is proportional to
$|F_{\alpha\beta}(\tau)|$. Hence, only if $|F_{\alpha\beta}(\tau)|\sim e^{-a\tau}\simeq e^{-b n}$ can the condition A2 be satisfied.
Introducing the  \emph{spectral density matrix of the bath} $R_{\alpha\beta}(\omega) =(1/2\pi)\int F_{\alpha\beta}(t)e^{-i\omega t} dt >0$ (Fourier transform\index{Fourier transform} of the correlation function,) one can use the Kubo-Martin-Schwinger (KMS) condition satisfied by all heat baths (we reintroduce $\hbar$ to indicate the quantum character of the KMS condition):
\begin{equation}
R_{\alpha\beta}(-\omega) = e^{-\hbar\omega/kT}R_{\beta\alpha}(\omega)\ .
\label{g1:KMS}
\end{equation}
Consider a single diagonal element $R_{\alpha\alpha}(\omega)$ corresponding to $F_{\alpha\alpha}(t)$.   Replace the actual model of a heat bath by a model with a simplified spectral density, still satisfying the KMS condition: 
\begin{equation}
R_{\alpha\alpha}(\omega) = R\  \mathrm{ if}\   \omega \geq 0 \ ,\  \mathrm{or}\ 
e^{\hbar\omega/kT}R  \ \mathrm{if}\  \omega < 0\ .
\label{g1:minbath}
\end{equation}
The asymptotic behavior of the autocorrelation function is then given by
\begin{equation}
|F_{\alpha\alpha}(t)|\simeq \frac{\hbar R}{kT}\frac{1}{t^2}, 
\label{g1:mincorr}
\end{equation}
and, obviously,  is not exponentially decaying. A similar $\sim 1/t^2$ tail is observed for a generic heat bath, because it is related to the jump $\hbar/kT$ of the first derivative  $R'_{\alpha\alpha}(\omega)$ at $\omega =0$, which is a consequence of the KMS condition.  This property of the thermal quantum autocorrelation function is often attributed to
the \emph{thermal memory} time $ \hbar/kT$. The presence of this nonexponential tail leads to the substantial  contribution of many-qubit errors. These considerations illustrate the challenges faced in applying versions of the threshold theorem\index{theorem!threshold} that rely on the ``locality assumption'' A2.

\subsection{Generalized threshold theorems and generic environments}
There exist attempts to prove threshold theorems under weaker assumptions, starting with Hamiltonian models and avoiding the unrealistic assumption A2 \cite{Terhal:2005:012336,Novais:2008:012314}. Before discussing these models, one should stress the fundamental assumption which should be imposed on the dynamics of the bath and the interaction
Hamiltonian in order to make the problem of FTQC non-trivial, and which was already briefly introduced in the previous section. In the phenomenological approach, it is assumed that the error per gate is fixed and cannot be scaled with the size of the computer. The corresponding assumption in the Hamiltonian approach with the interaction Hamiltonian of the form (\ref{g1:int})
is that the spectral density matrix\index{spectral density} for the reservoir
\begin{equation}
R_{\alpha\beta}(\omega) =\frac{1}{2\pi}\int_{-\infty}^{\infty} e^{-i\omega t} \mathrm{Tr}\bigr(\rho_B B_{\alpha}(t) B_{\beta}\bigr) dt 
\label{g1:auto}
\end{equation}
satisfies two conditions (for $\omega\in[0, \Omega]$):

\par
R1)
\begin{equation}
{\cal R}(\omega)-R(\omega)I \geq 0
\label{g1:R1}
\end{equation}
where the matrix ${\cal R}(\omega)$ has matrix elements $R_{\alpha\beta}(\omega)$.
\par
R2)
\begin{equation}
R(\omega)\geq \omega \eta \ , \quad \eta > 0\ .
\label{g1:R2}
\end{equation}

The first condition means that, for a fixed $\omega$, the eigenvalues of ${\cal R}(\omega)$ are bounded from below by $R(\omega)$. On the other hand, those eigenvalues  describe the dissipation rates for the degrees of freedom of the open system which oscillate with the frequency $\omega$. This relation will be better understood in the Markovian limit (see the next Section). A simple argument can involve the\index{Fermi Golden Rule} Fermi Golden Rule, which associates the transition probabilities, and hence dissipation (decoherence) rates, to the effective density of the bath's excitations at the energy $E= \hbar\omega$, strictly related to the spectral density. Therefore, the condition R1 eliminates the situations where a certain system's degrees of freedom do not dissipate (or decohere) at all. A particular example are decoherence-free subspaces \cite{Lidar:1998:2594} generated by a collective coupling to a single bath of the form 
\begin{equation}
H_{int} = \sum_{\mu = x,y,z}(\sum_{j=1}^N\sigma^{\mu}_j)\otimes B_{\mu}.
\label{g1:coll}
\end{equation}
Indeed, the collective coupling (\ref{g1:coll}) produces a correlation matrix of the tensor form $R_{\mu j,\nu k}(\omega)= R_{\mu\nu}(\omega)$ which has all but three eigenvalues equal to zero.\\ 

The condition R1 means that the relaxation rates of the original model are larger then the relaxation rates of the simplified one with diagonal correlation matrix $R(\omega)\delta_{\alpha\beta}$. The latter corresponds to a model with identical ``private baths" coupled to each $\sigma^{\alpha}$.  Such a simplified model shows at most slower relaxation to the equilibrium state, and is very useful if we want to estimate the slowest relaxation time of the system from above. 
\par
The condition R2 corresponds to the ``fixed error $\eta$ for a single gate'' assumption in the phenomenological approach. To show this, one can use the Margolus-Levitin theorem\index{theorem!Margolus-Levitin} \cite{Margolus:1998:188}:
\emph{A quantum system of energy E needs at least a time $\simeq\frac{\hbar}{ E}= \frac{1}{\omega}$ to go from one state to an orthogonal state}.
On the other hand, as argued above,  $R(\omega)$ can be seen as the lower bound for the relaxation rate of 
the modes with  frequency $\omega$ (i.e., corresponding to the energy difference $E=\hbar\omega$). As a consequence, a typical gate needs a time of the order of $\tau\simeq\frac{\hbar}{ E}= \frac{1}{\omega}$, and hence the error due to the interaction with a bath during the gate's execution is roughly bounded from below by $\tau R(\omega)= R(\omega)/\omega$.\\
Note that a commonly used anzatz for the spectral density, of the form
\begin{equation}
R(\omega)= C \omega^d e^{-\omega/\Omega}, \quad \mathrm{with}\quad d \geq 1,
\label{g1:Anz}
\end{equation}
does not satisfy the condition R2. For such a coupling one can effectively eliminate errors by appropriately scaling down the energy used to implement gates \cite{Alicki:2002:062101}. On the other hand, such a spectral density is typical for linear coupling to bosonic heat baths (photons, phonons, etc.). Unfortunately, there  always exist other mechanisms, such as elastic scattering, which lead to a finite dephasing rate characterized by the value $R(0) > 0$.
\par
We now discuss briefly two Hamiltonian models of FTQC. In the  non-Markovian model of  Terhal and Burkard \cite{Terhal:2005:012336}, the assumption of ``small norm'' of the interaction Hamiltonian implies  that the operator norm (largest eigenvalue) of each of the bath operators $B$ in the interaction Hamiltonian (\ref{g1:int}) satisfies
\begin{equation}
\|B\|_{\infty}\tau \leq\epsilon, 
\label{g1:small}
\end{equation}
where $\epsilon \ll$1 is a dimensionless small constant characterizing the decoherence\index{decoherence} strength, and where $\tau$ is the execution time of a logic gate.\index{logic gates}
Adding the definition of the spectral density and the condition R2, together with $\omega \in [0,\Omega]$, one obtains a sequence of inequalities:
\begin{equation}
\frac{1}{2}\eta \Omega^2  < \int_{-\infty}^{\infty}R(\omega) d\omega = \mathrm{Tr}(\rho_B B^2)\ \leq \|B\|_{\infty}^2 \leq\epsilon^2 \tau^{-2}\ .
\label{g1:snorm}
\end{equation}
This implies a cutoff-dependent bound on the gate time $\tau$. As the physics should not depend strongly on the particular value of the cut off frequency $\Omega$ the small norm assumption is difficult to defend.

The second example of the Hamiltonian modeling is presented in \cite{Novais:2008:012314}  where the ideas of renormalization group\index{renormalization group} techniques are applied. The arguments are not fully rigorous, certain simplifying hypotheses concerning the computer-bath interaction are used without proofs. An example of  such a condition is  \emph{the ``hypercube'' assumption},\index{hypercube} which is
difficult to justify, as the correlations between neighboring qubits are due to the same interactions that are needed to couple the qubits in the process of error correction. There are a number of other delicate points that one must be sure to treat carefully. For example, the authors of \cite{Terhal:2005:012336} and \cite{Novais:2008:012314} use the basic ingredients of  the theory of fault tolerance, including, e.g., the constant supply of \emph{fresh qubits}, and the assumption that error propagation\index{error!propagation in fault-tolerant quantum circuits} is handled by the quantum code. However, these ideas need a rigorous first-principles background.
Not all FTQC analyses fully conform to these requirements.

\section{Fault-tolerance and Quantum Memory}

The previous discussion demonstrates that a rigorous proof of the validity of quantum fault-tolerance, based on a first principles Hamiltonian analysis, is far from being complete. The main problem is related to the new time scale introduced by the external control, which is not well separated from the other time scales of the problem such as, e.g., thermal memory time and the inverse of the cut-off frequency for the bath, or relaxation time scales for the computer. Therefore, standard approximation techniques, such as the  Markovian  or adiabatic limits, cannot be directly applied \cite{Alicki:2006:052311}. Moreover, the problem of fault-tolerance belongs to the category of \emph{subtle problems}, in the sense that even reasonable but not rigorously controlled approximations can produce completely false results (compare mean-field approximation in the theory of phase transitions). As a consequence, it is prudent first to try to solve rigorously a simpler problem: the existence (or perhaps non-existence) of a stable quantum memory.  Any quantum computer could be used as a quantum memory, and the preservation of an arbitrary state of an encoded qubit for a long time can be treated as the simplest quantum algorithm. 
\par
One can consider two cases of quantum memory: a dynamical one, based on the standard model of a quantum computer with unitary gates, ancillas, etc.,
and a self-correcting one, with a properly designed time-independent Hamiltonian that protects certain degrees of freedom. Actually, these two cases  should be equivalent from both the physical and mathematical point of view. The first type of memory is described by time-periodic Hamiltonians, that correspond to state recovery cycles by error correcting procedures. However, periodic Hamiltonians are mathematically very similar to time-independent ones \cite{Cycon:1987:Springer}, and in the theory of open systems there exists a construction of Markovian dynamics\index{Markovian dynamics} for periodic Hamiltonians which is very similar to the derivation of the Davies generators used for self-correcting model Hamiltonians \cite{Alicki:2006:052311}. Therefore, in the following only self-correcting models of quantum memory\index{quantum memory} will be considered.

\subsection{Definition of quantum memory}

A many body quantum system consisting of $N$ elementary subsystems (e.g., qubits), and hence described by an algebra of bounded operators ${\cal A}_N$  and a Hamiltonian $ H_N$, provides a model of a \emph{scalable quantum memory}
if there exists at least  one pair of Hermitian operators  $X, Z \in {\cal A}_N$ corresponding to an encoded robust qubit satisfying the following conditions:

M1) They generate a qubit algebra, i.e., $X^2 = Z^2 = 1$, $XZ + ZX = 0$.

M2) They are physically implementable i.e. one can construct perturbed Hamiltonians of the form
\begin{equation}
H_N(t)  = H_N + f_x(t)X + f_y(t)Y + f_z(t)Z
\label{g1:perham}
\end{equation} 
where $\{f_i\}$ represent  external classical fields which allow control over the qubit, and $Y= iZX$.

M3) They are stable with respect to thermal noise. This  can be described in terms of autocorrelation function decay     
\begin{equation}
|\langle X(t) X\rangle_{eq}| \sim e^{-\gamma_N t}
\label{g1:auto1}
\end{equation} 
and similarly for $Z$. The decay rate should satisfy
\begin{equation}
\lim_{N\to\infty} \gamma_N = 0\ ,
\label{g1:decrate}
\end{equation} 
preferably exponentially fast (\emph{exponentially stable memory}). 
\par
The average in (\ref{g1:auto1}) is taken with respect to the thermal equilibrium state of the total system consisting of our candidate for the memory and a heat bath. The Heisenberg evolution of $X(t)$, etc., is also governed by the total Hamiltonian of the system and bath in the weak coupling regime. Note that the decay of autocorrelation functions (\ref{g1:auto1}) implies a similar decay of the averaged state's fidelity, which is a common measure characterizing the quality of quantum information stored in a noisy environment \cite{Alicki:2009:012316}. 

\subsection{Markovian model of self-correcting quantum memory}

\index{self-correcting quantum memory}
Following \cite{Alicki:2007:Springer}, consider the scheme presented in Section 1.2.1 but with a constant bare Hamiltonian ${H_Q}'$ and a system-bath interaction Hamiltonian of the form 
\begin{equation}
H_{int} = \lambda \sum_{\alpha} S_{\alpha}\otimes B_{\alpha} ,
\label{g1:hamint}
\end{equation}
with an explicitly small coupling constant $\lambda$, and $S_{\alpha}$ denoting, for example, $\sigma_j^{x,y,z}$ from Eq. (\ref{g1:int}). Denote by $\{\omega \}$ the set of eigenfrequencies
of the renormalized, physical Hamiltonian $H_Q$,
and let $S_{\alpha}({\omega })$ be the discrete Fourier components of $S_{\alpha}$ in the
interaction picture, i.e., 
\begin{equation}
S_{\alpha}(t)=\exp (iH_{Q}t)S_{\alpha}\exp (-iH_{Q}t)=\sum_{\{\omega\} }S_{\alpha}({\omega })\exp (i\omega
t). 
\label{g1:eq:S}
\end{equation}
According to the nontriviality condition (\ref{g1:R1}), it suffices to consider models with \emph{private baths}, i.e., independent,
identical heat baths for each degree of freedom corresponding to $S_{\alpha}$.
A sequence of approximations, discussed for example in \cite{Alicki:2006:052311}, leads to the following 
Markovian master equation\index{master equation!Lindblad} of the Lindblad-Gorini-Kossakowski-Sudarshan\index{Lindblad-Gorini-Kossakowski-Sudarshan master equation} \cite{Lindblad:1976:119,Gorini:1976:821} type, derived rigorously
in terms of van Hove \emph{weak coupling limit}\index{weak coupling limit}\index{van Hove limit} by Davies \cite{Davies:1974:91}:
\begin{equation}
\frac{d\rho}{dt} = -i[H_{Q},\rho ]+\mathcal{L}\rho , 
\end{equation}
\begin{equation}
\mathcal{L}\rho \equiv \frac{1}{2}\lambda ^{2}\sum_{\alpha}\sum_{\{\omega \}}R(\omega)
\Bigl([S_{\alpha}(\omega ),\rho S_{\alpha}(\omega )^{\dagger }]+[S_{\alpha}(\omega )\rho ,S_{\alpha}(\omega)^{\dagger }]\Bigr), 
\label{g1:Dav}
\end{equation}
with the spectral density satisfying the KMS condition
$R(-\omega)= e^{-\omega/kT} R(\omega)$.

It is convenient to use the Heisenberg picture version of the evolution Eq. (\ref{g1:Dav})
\begin{equation}
\frac{dA }{dt} = i\mathcal{H}A+\mathcal{L^*}A, \quad  \mathcal{H}A \equiv [H_{Q},A], 
\end{equation}
\begin{equation}
\mathcal{L}^*A \equiv \frac{1}{2}\lambda ^2\sum_{\alpha}\sum_{\{\omega \}}R(\omega)
\bigl(S_{\alpha}(\omega)^{\dagger }[A , S_{\alpha}(\omega )]+[S_{\alpha}(\omega)^{\dagger}, A]S_{\alpha}(\omega )\bigr). 
\label{g1:DavH}
\end{equation}

The sum  $\mathcal{G}= i\mathcal{H} + \mathcal{L^*}$ generates a semi-group of completely positive, identity preserving transformations on the algebra of observables. However, due to its specific form, it enjoys a number of important additional properties \cite{Alicki:2007:Springer}: \\

D1) The canonical Gibbs state\index{Gibbs!state} is stationary
\begin{equation}
\mathrm{Tr} \Bigl( \rho_\beta\,  e^{t\mathcal G}(X) \Bigr) =  \mathrm{Tr}\bigl( \rho_\beta\, X \bigr) 
\label{g1:D1}
\end{equation}
with
\begin{equation}
\rho_\beta = \frac{ e^{-\beta H_{Q}}}{\mathrm{Tr} \Bigl( e^{-\beta H_{Q}}\Bigr)},
\label{g1:invstate}
\end{equation}
D2) The semi-group is relaxing: any initial state $\rho$ evolves to $\rho_\beta$
\begin{equation}
\lim_{t\to\infty}\ \mathrm{Tr} \Bigl( \rho\,  e^{t\mathcal G}(X) \Bigr) = \mathrm{Tr} \bigl( \rho_\beta\, X \bigr),
\label{g1:relax}
\end{equation}
D3) $\mathcal{L}^*$  satisfies the detailed balance condition, often called reversibility
\begin{equation}
\mathcal{H}\mathcal{L}^*=\mathcal{L}^*\mathcal{H}
\label{g1:D2}
\end{equation}
and
\begin{equation}
\mathrm{Tr}\Bigl(\rho_\beta\, Y^\dagger\, \mathcal{L}^*(X) \Bigr) =  \mathrm{Tr}\Bigl(\rho_\beta\, \bigl(\mathcal {L}^*(Y)\bigr)^\dagger\, X \Bigr).
\label{g1:db}
\end{equation}
Equation~(\ref{g1:db}) expresses the self-adjointness of $\mathcal{L}^*$ with respect to the Liouville scalar\index{scalar!Liouville product} product
\begin{equation}
\langle X \,,\, Y\rangle_\beta := \mathrm{Tr}\rho_\beta\, X^\dagger\, Y.
\label{g1:lio}
\end{equation}
D4) The dissipative part $\mathcal{L}^*$ of the generator is negative definite. 
\par
Due to D2 any initial state of a system will eventually relax to equilibrium. Information can be encoded by perturbing the equilibrium state of the system and, in order to retrieve this information, one must single out observables that detect the perturbation of the state. To encode a single qubit one needs metastable observables $X,Y,Z$ satisfying conditions M1-M3 of the previous section. It is natural to search for such observables among the constants of motion for the Hamiltonian that have zero expectation values in the Gibbs state:\index{Gibbs!state}
\begin{equation}
\langle R \rangle_\beta = 0.
\label{g1:bias}
\end{equation}
Hence, for the Markovian model above, the following estimation holds ($R= X,Y,Z$):
\begin{equation}
\langle R \,,\,e^{t\mathcal{G}}R\rangle_\beta =\langle R \,,\,e^{t\mathcal{L}^*}R\rangle_\beta \geq  \exp\{t\langle R \,,\,\mathcal{L}^*R\rangle_\beta\}.
\label{g1:core}
\end{equation}
One proves it easily, by decomposing $R$ into 
normalized eigenvectors  of $\mathcal{L}^*$, and using convexity of the function 
$e^{-x}$.
\par
It follows from Eq.~(\ref{g1:core}) that the necessary condition for the existence of such (exponentially) metastable observables\index{metastable observable} is exponentially fast vanishing of the  matrix elements $\langle R \,,\,\mathcal{L}^*R\rangle_\beta$. In particular, one can expect the following scaling:
\begin{equation}
|\langle R \,,\,\mathcal{L}^*R\rangle_\beta |\sim e^{-cN^p},
\label{g1:core1}
\end{equation}
with constants $c,p >0$ independent of $N$.
As $R$ is orthogonal to $I$ (the nondegenerate eigenvector of $\mathcal{L}^*$ with eigenvalue $0$), and $\|R\|_{\beta} =1$, the matrix element
 $\langle R \,,\,\mathcal{-L}^*R\rangle_\beta$ is bounded from below by a spectral gap for
the self-adjoint operator $-\mathcal{L}^*$ (i.e., its lowest eigenvalue is different from  $0$). It shows that the stability analysis of encoded qubits
relies on the investigation of the spectrum of $-\mathcal{L}^*$ in the neighborhood of zero.
\par
Finally, one should also remember the condition M2 which requires that the encoded qubit observables must be efficiently implementable.

\subsection{Kitaev models}

The family of Kitaev models in\index{Kitaev models} $D=2,3,4$ dimensions \cite{Kitaev:2003:2,Dennis:2002:4452} consists of spin-1/2 models on a $D$-dimensional lattice with a\index{toric topology} toric topology, and with a Hamiltonian exhibiting the special structure:
\begin{equation}
H = -\sum_{s} X_s - \sum _{c} Z_c .
\label{g1:KitH}
\end{equation}
Here, $X_s = \otimes_{j\in s}\sigma^x_j$, $Z_c = \otimes_{j\in c}\sigma^z_j$ are products of Pauli matrices belonging to certain finite sets on the lattice: ``stars'' and ``cubes.''\index{lattice!for Kitaev models} They are chosen in such a way that all $X_s , Z_c$ commute and form an Abelian subalgebra $\mathcal{A}_{ab}$ in the total algebra of $2^N\times 2^N$ matrices. The commutant of  $\mathcal{A}_{ab}$, denoted by $\mathcal{C}$, is noncommutative, and provides a natural candidate for  the subalgebra containing encoded qubit observables. Indeed, for all $D=2,3,4$ one can define \emph{bare qubit observables}  $X^{\mu}, Z^{\mu}\in\mathcal{C}$ where  $\mu = 2,3,4$ correspond to $D$ independent encoded qubits. They are products of the corresponding Pauli matrices over topologically nontrivial loops (surfaces). The choice of loops is, of course, not unique.

To discuss the question of stability with respect to thermal noise one can use the Markovian models with Davies generators described in the previous section. In this context, Kitaev models are particularly simple \cite{Alicki:2007:6451}. The commutation of all Hamiltonian terms implies a strict\index{locality} locality
 of the model (absence of wave propagation), and implies that the Fourier components in (\ref{g1:eq:S}) are local and correspond to only a few Bohr frequencies,\index{Bohr frequency} independent of the size of the system. This makes the analysis of spectral properties of the Davies generator feasible. Despite this simplification, the proofs of the results are too involved to be reproduced here; we refer the reader to \cite{Alicki:2007:6451,Alicki:2009:065303} for details. 
\par
For the 2D-Kitaev model\index{Kitaev model!two-dimensional} it is enough to take the terms containing $\sigma^x, \sigma^z$ in the interaction Hamiltonian (\ref{g1:int}). Then the form of the Markovian master equation in the Heisenberg picture is the following\index{master equation!Heisenberg picture} \cite{Alicki:2007:6451}:
\begin{equation}
 \frac{dA}{dt}
 = i[H,A] 
  + \frac{1}{2} \sum_{j=1}^N \Bigl\{  \Bigl(
 a_j^\dagger\, [A,a_j] + [a_j^\dagger,A]\, a_j +  e^{-2\beta}\, a_j\,
 [A,a_j^\dagger] + e^{-2\beta}\, [a_j,A]\, a_j^\dagger \Bigr) 
 - [a_j^0, [a_j^0,A]] \Bigr\} 
 \end{equation}
 \begin{equation}
 + \frac{1}{2} \sum_{j=1}^N \Bigl\{ \Bigl(
 b_j^\dagger\, [A,b_j] + [b_j^\dagger,A]\, b_j +  e^{-2\beta}\, b_j\,
 [A,b_j^\dagger] +  e^{-2\beta}\, [b_j,A]\, b_j^\dagger \Bigr) 
- [b_j^0, [b_j^0,X]] \Bigr\}.
\label{g1:MME} 
\end{equation}
We do not define the operators $a_j, a_j^0, b_j, b_j^0$ here, but rather give their physical interpretation. The operator $a_j$ ($a_j^\dagger$) annihilates
(creates) a pair of excitations (or anyons)\index{anyon}attached to the site $j$, and corresponding to the part of the Hamiltonian $-\sum Z_c$ in (\ref{g1:KitH}) (type-$Z$ anyons), while 
$a_j^0$ generates diffusion of anyons of the same type. Similarly, the operators $ b_j, b_j^\dagger , b_j^0$ correspond to the type-$X$ anyons.
From the form of the Hamiltonian, it follows that the 2D-Kitaev model is equivalent to a gas of noninteracting particles (anyons of two types), which are created/annihilated in pairs, and  diffuse. Hence, heuristically, no mechanism of macroscopic free energy barrier between different phases
is present that could be used to protect even classical information. Mathematically, it was proved that the dissipative part of the Davies generator possesses a spectral gap independent of the size $N$, and therefore no metastable observables exist in this system \cite{Alicki:2009:065303}. The main tool used in the proof is the fact that for a positive operator $K$ acting on the Hilbert space ,  the inequality $K^2 \geq c K$, $c>0$ implies that  the spectral gap of $K$ is bounded from below by the number $c$.  Another useful property of the Davies generator is that it is a sum of many negatively defined terms. Hence, skipping some of them can simplify estimations without increasing the spectral gap.
\par
The 4D-Kitaev model\index{Kitaev model!four-dimensional} is much more interesting. Instead of noninteracting particles, a picture similar to droplets in the 2D-Ising model\index{Ising model} appears \cite{Dennis:2002:4452}. The excitations of the system are represented by closed loops, with energy proportional to the loops' lengths. This provides the sought-after mechanism
of a macroscopic energy barrier separating topologically nonequivalent spin configurations. The 3D-Kitaev model\index{Kitaev model!three-dimensional} provides this mechanism for one type of excitations only. Therefore, only the encoded ``bit'' is protected, but not the ``phase." The structure of the evolution equation is always similar to (\ref{g1:MME}), with the operators $a_j^{\dagger}, b_j^{\dagger}$ creating excitations of two types, and $a_j^0, b_j^0$ changing the shape of excitations but not their energy. It seems necessary to use the full interaction Hamiltonian (\ref{g1:int}), which leads to additional processes of energy transfer between the two types of anyons\index{anyon}.
\par
In the paper \cite{Alicki:2010:1} it is proved that, for the 4D model, there exist exponentially  metastable \emph{dressed} qubit observables ${\tilde X}^{\mu}, {\tilde Z}^{\mu}\in\mathcal{C}$ with  $\mu = 1,2,3,4$, related to the bare ones by the formulas
\begin{equation}
 {\tilde X}^{\mu} =X^{\mu}F^{\mu}_x ,\quad {\tilde Z}^{\mu} =Z^{\mu}F^{\mu}_z,
\label{g1:qbitobs}
\end{equation}
where $F^{\mu}_z , F^{\mu}_z$ are Hermitian elements of the algebra $\mathcal{A}_{ab}$ with eigenvalues $\pm 1$. On the other hand, bare qubit observables are highly unstable, with relaxation times $\sim \sqrt{N}$. The metastability of (\ref{g1:qbitobs})
is proved using the Peierls argument\index{Peierls argument} applied to classical ``submodels'' of the 4D-Kitaev model\index{Kitaev model!four-dimensional} generated either by $-\sum_{s} X_s$ or $ - \sum _{c} Z_c $.
The main mathematical tool is the following inequality \cite{Alicki:2010:1}:
\begin{equation}
-\langle A,{\cal L}^*A)\rangle_\beta \leq 2  \max_{\{\omega\}}\{R(\omega)\} \sum_\alpha 
\langle[S_\alpha, A],[S_\alpha, A]\rangle_\beta ,
\label{g1:eq:przerwa}
\end{equation}
valid for any  Davies generator\index{Davies generator} (\ref{g1:Dav}) and any $A$ in the eigenspace of $[H,\cdot]$. The advantage of this formula is the absence of Fourier components $S_{\alpha}(\omega)$, replaced now by much simpler $S_{\alpha}$.

The metastable observable\index{observable!metastable} (say ${\tilde X}^{\mu}$) is constructed by the following operational procedure, which determines its outcomes:

1. Perform a measurement of all observables $\sigma^x_j$.

2. Compute the value of the bare observable $X^{\mu}$ (multiply previous outcomes for spins belonging to the ''surface'' which defines $X^{\mu}$).

3. Perform a certain classical algorithm (polynomial in $N$)  which allows to compute from the $\sigma^x_j$- measurement data the value $\pm 1$ of ''correction'', i.e., the eigenvalue of
$F^{\mu}_x$.

4. Multiply the bare value by the correction to get the outcome of ${\tilde X}^{\mu}$.

Although the observables defined by the operational procedure of above satisfy M1 and M3, the condition M2 appears problematic. It is hard to imagine any efficient construction of the corresponding operators that could be used to design the control Hamiltonians (\ref{g1:perham}). The measurement
of individual spins which is necessary for the extraction of the ${\tilde X}^{\mu}$'s outcome is destructive and not repeatable, at least for the model with the full interaction Hamiltonian (\ref{g1:int}). Therefore, at present the 4D-Kitaev model seems problematic even as a classical memory.

An attempt to solve the problems for a quantum memory is presented in \cite{Bombin:0907.5228}, where 6D topological color codes\index{code!color} are discussed. Those systems admit both the stable encoding of qubits based on the similar mechanism to 4D-Kitaev model, and local transversal unitary gates. However, the gate Hamiltonians do not commute with the protecting
Hamiltonian, and therefore the effective map acting on the encoded qubits is dissipative. To avoid this phenomenon, the authors proposed to switch off a part of protective Hamiltonian, but the consequences of this procedure were not discussed rigorously. 
\section{Concluding remarks}
As shown in the chapter, the generic non-exponential decay of thermal autocorrelation functions and the 4D-Kitaev model, illustrate the serious difficulties associated with the idea of FTQC. For the quantum computer Hamiltonian model, the lower the temperature  the more correlated the noise (``thermal memory''). For the Kitaev model, the better the protection of the encoded qubit observables, the more difficult is their control and accessibility. It is quite plausible that these difficulties are fundamental and not technical, and that a kind of ``Heisenberg relation'' is at work. This could be related to the fact that the same physical interactions used to control a system provide the coupling of that system to an environment. 

Another observation valid for topological memories is that the mechanism of information protection is essentially the same for the classical and quantum cases:  the existence of free energy barriers separating metastable states. It seems that to perform a gate one has to overcome such a barrier, which involves a suitable amount of work that is then dissipated into the environment. Therefore, it is plausible to expect that there exists a fundamental conflict between  stability of the encoded information and reversibility (in the sense of Hamiltonian, non-dissipative dynamics) of its processing \cite{Alicki:2013:4910}. It does not harm classical information processing, as it can be and is, in all practical implementations, performed by strongly dissipative dynamical maps; but quantum information based on quantum coherence needs unitary (Hamiltonian) gates.\\
We are still far from  a quantitative understanding of these relations and associated bounds,  and further studies are necessary to clarify these questions. It is quite possible that the ultimate bounds on the efficiency of quantum information processing will be provided by phenomenological thermodynamics, in particular by its Second Law \cite{Alicki:2009.0811}.

\end{document}